\documentclass[]{aa}
\usepackage{graphics,latexsym,psfig}
\usepackage{graphicx}

\def\farcs{\hbox{$.\!\!^{\prime\prime}$}}  % Fractions of arcseconds
\def\fdegs{\hbox{$.\!\!^{\circ}$}}  % Fractions of degrees
\def\asec{\ifmmode ^{\prime\prime}\else$^{\prime\prime}$\fi}
\def\amin{\ifmmode ^{\prime}\else$^{\prime}$\fi}
\def\degs{\ifmmode ^{\circ}\else$^{\circ}$\fi}
\def\etal{{et\,al. }}
\def\msun{M$_{\odot}$}

\begin{document}

\title{The host of GRB/XRF 030528 - an actively  star forming galaxy at z=0.782\thanks{Based on observations
       collected at  the European Southern Observatory, La  Silla and Paranal,
       Chile under ESO Program 075.D-0539.}  }

\titlerunning{The host galaxy of GRB/XRF 030528}

\author{
A. Rau \inst{1}        \and
M. Salvato\inst{1} \and
J. Greiner \inst{1} 
}

\offprints{A. Rau, arau@mpe.mpg.de}

\institute{Max-Planck-Institut f\"ur extraterrestrische Physik,
  Giessenbachstrasse, 85748 Garching, Germany 
}

% --------------------------------------------------------------------------
\date{Received 6 July  2005 / Accepted 17 August 2005}

\abstract{An important  parameter for the distinction of  X-ray flashes, X-ray
rich bursts  and Gamma-ray  bursts in the  rest frame  is the distance  to the
explosion site.  Here we report on the spectroscopic redshift determination of
the host galaxy of XRF/GRB 030528 using the ESO VLT FORS2 instrument. From the
strong  oxygen and hydrogen  emission lines  the redshift  was measured  to be
$z$=0.782$\pm$0.001.  Obtaining the line  luminosities and ratios we find that
the  host  is consistent  with  being an  actively  star  forming galaxy  with
sub-solar  metallicity.  With  a stellar  mass of  $\sim$10$^{10}$\,\msun\ the
host is placed among the most massive GRB host galaxies at a similar redshift.
Estimating  the redshifted  properties of  the prompt  emission, we  find that
XRF/GRB 030528 would  be classified as an X-ray rich bursts  in the rest frame
rather than an X-ray flash in the typically used observer frame.
} \maketitle

% --------------------------------------------------------------------------
\section{Introduction}

The  distribution   of  observer   frame  peak  energies   in  $\nu$F$_{\nu}$,
$E^{obs}_{peak}$,  of  the prompt  high-energy  emission  of Gamma-ray  bursts
(GRBs)  ranges from  several keV  up to  a few  MeV with  a  clustering around
250\,keV (Preece  \etal 2000).  Recently,  Special attention was drawn  to the
extreme  low  energy part  of  the  distribution.   Based on  {\it  Beppo-SAX}
Wide-Field Camera  observations Heise \etal (2001) classified  these events as
so-called X-ray  flashes (XRFs).   XRFs are similar  to long duration  GRBs in
various  prompt  burst  properties  (e.g.    the  duration  and  the  low  and
high-energy     spectral     slopes)      but     are     characterized     by
$E^{obs}_{peak}$$\le$30\,keV.   An  operational  classification based  on  the
ratio         of         X-ray         to        $\gamma$-ray         fluence,
log($S_X$(2--30\,keV)/$S_\gamma$(30--400\,keV))  in  the  observer frame,  was
proposed for  events detected  by the {\it  HETE-2} satellite  (Sakamoto \etal
2004).  According to this definition,  one third of all {\it HETE-2}-localized
bursts  are X-ray  flashes with  log($S_X$/$S_\gamma$)$>$0, another  third are
X-ray rich  bursts (XRR)  having log($S_X$/$S_\gamma$)$>$--0.5, with  the rest
being the ``classical'' GRBs (Lamb \etal 2005).

It has been suggested that XRFs represent the extension of GRBs to bursts with
low peak energies  and that XRFs, XRRs and GRBs form  a continuum (Heise \etal
2001, Barraud \etal 2003).  This  was strengthened by the discoveries of X-ray
(Harrison \etal 2001), optical (Soderberg  \etal 2003) and radio (Taylor \etal
2001) afterglows  for some  XRFs with  properties similar  to those  found for
GRBs. Furthermore, for XRF 020903 (Soderberg \etal 2005) and XRF 030723 (Fynbo
\etal 2004) possible  supernova bumps in the afterglow  light curves have been
reported similar to  the observed associations of the  deaths of massive stars
with long-duration GRBs (e.g.  Hjorth \etal 2003; Stanek \etal 2003; Zeh \etal
2004).

A variety of theoretical models has been proposed to explain the observed peak
energies of XRFs.  E.g., (i) a high  baryon loading in the GRB jets can result
in bulk Lorentz factors much smaller than those expected in GRBs (Dermer \etal
1999;  Huang \etal  2002)  (ii) Similarly,  a  low contrast  between the  bulk
Lorentz  factors of  the colliding  relativistic shells  can  produce XRF-like
events  (Barraud  \etal 2005).   (iii)  For GRB  jets  which  are not  pointed
directly at the  observer the spectrum will be softer  as well (Yamazaki \etal
2002). (iv) The peak energies of GRBs  at high redshift will be moved to lower
energies and can mimic XRFs in the observer frame (Heise \etal 2001).

An important discriminator between the individual models is the distance scale
to  the explosion  site. A  number of  XRFs have  been localized  to arcminute
accuracy in the past but only  for few the accurate distance could be obtained
so far. For XRF 020903 and  030429 the redshift could be measured unambigously
as  $z$=0.251 (Soderberg  \etal  2004) and  $z$=2.65  (Weidinger \etal  2003),
respectively.   A third  event, XRF  040701, has  a candidate  host  galaxy at
$z$=0.2146  (Kelson \etal  2004) associated  with one  of the  two tentatively
fading X-ray  sources in the  error box (Fox  2004). For some more  XRFs upper
limits on  the redshift could be  set from optical  follow-up observations. In
two cases, XRF 011030 and 020427, host galaxies with typical properties of GRB
hosts have been detected and photometric evidence suggests the redshifts to be
$z$$<$3.5  (Bloom  \etal 2003)  and  $z$$<$2.3  (van  Dokkum \&  Bloom  2003),
respectively.  For  XRF 030727 a firm  upper limit of $z$=2.3  could be placed
from the absence of Ly$\alpha$ absorption and prominent lines in the afterglow
spectrum  as  well  as  from  a  light curve  bump  associated  with  possible
underlying supernova (Fynbo \etal 2004).

Here  we  report on  the  spectral  analysis of  the  host  galaxy of  GRB/XRF
030528. After summarizing  the known properties  of the burst and  host (Sect.
2) we present  the observations  and data reduction  (Sect.  3).  The obtained
host  properties derived  from the  spectrum  are shown  in Sect.   4 and  the
implications in the context of XRFs and GRB/XRF host galaxies are discussed in
Sect. 5.

We used a cosmology  of $\Omega_m$=0.27, $\Omega_\lambda$=0.7 and H$_0$=70\,km
s$^{-1}$  Mpc$^{-1}$ throughout  the  paper.  All  photometric magnitudes  are
given in the AB system.

\section{GRB/XRF 030528}

The  high energy transient  was detected  by {\it  HETE-2} as  a long-duration
Gamma-ray                  burst                 (HETE                 trigger
\#2724)\footnote{http://space.mit.edu/HETE/Bursts/Data}.     The   event   was
moderately  bright  with  a  fluence of  $S$=5.6$\pm$0.7$\times$10$^{-6}$  erg
cm$^{-2}$ and a  peak flux on a one second  time scale of 4.9$\times$10$^{-8}$
erg cm$^{-2}$  s$^{-1}$ in  the 30-400\,keV band  (Sakamoto \etal  2005).  The
burst duration (given as T$_{90}$, which  is the time over which a burst emits
from 5\%  of its total  measured counts to 95\%)  was T$_{90}$=49.2$\pm$1.2\,s
(30-400\,keV)  and the  high energy  spectrum peaked  at  32$\pm$5\,keV.  With
log($S_X$/$S_\gamma$)=0.04,  the  event  was  classified as  an  X-ray  flash.
Accordingly, we will  refer to the event as XRF  030528 throughout this paper,
while at other places (e.g.  Butler \etal 2004; Rau \etal 2004, R04 hereafter)
the identifier GRB 030528 was used.

The  burst occured  in a  crowded field  in the  sky near  the  Galactic Plane
(LII=0\fdegs0462 \& BII=11\fdegs2902) which lead to complication of the ground
based  optical follow-up  observations  by a  significant Galactic  foreground
extinction.  A faint  near-IR afterglow was detected (Greiner  \etal 2003) and
at  the same  position a  fading  X-ray source  was found  with {\it  Chandra}
(Butler \etal 2003).

In  R04 we described  the properties  of the  near-IR afterglow  as well  as a
photometric  study of  the detected  underlying host  galaxy.  The  galaxy was
found    to   be   among    the   brightests    observed   GRB    hosts   with
$K_{AB}$$\sim$21.8$\pm$0.7\,mag.   A  fitting   of  template  spectral  energy
distributions (SEDs) to the photometry in $V,R,I,J_s,H$ \& $K$ showed that the
host  properties were consistent  with that  of a  young star  forming galaxy.
Unfortunately, the lack of spectroscopy and the sparse photometric sampling of
the  SED did  not allow  to determine  the redshift  accurately, but  the data
favored $z$$<$1.

% --------------------------------------------------------------------------
\section{Observations and Data Reduction}

The  host  galaxy of  XRF  030528  was observed  with  the  Focal Reducer  and
low-dispersion Spectrograph 2  (FORS2) at the 8.2\,m ESO  Very Large Telescope
(VLT) Antu in Paranal, Chile. A  total of twelve exposures, each lasting 594s,
were taken  in two nights,  on Apr.   12, 2005 and  May 6, 2005.   We obtained
longslit spectroscopy using the 300V  grism together with the order separation
filter GG435, thus covering a spectral range of approx. 5200--9200\,\AA. Using
a 1\farcs0  slit, the  3.3\,\AA\ pixel$^{-1}$ scale  leads to a  resolution of
13.5\,\AA (FWHM) at 1\farcs0 seeing.

Flat-field, bias correction and cosmic ray removal was applied in the standard
fashion  using  IRAF\footnote{IRAF  is  distributed by  the  National  Optical
Astronomy Observatories, which are operated by the Association of Universities
for Research in Astronomy, Inc., under cooperative agreement with the National
Science Foundation.}.  The wavelength  calibration of the combined spectra was
done using HgCdHe+Ar  calibration lamps. The standard star  LTT 7379 (spectral
type  G0) was used  for the  flux calibration  of the  spectra. In  addition a
correction for telluric  absorption was performed using an  observation of the
telluric standard star EG 274 (spectral type DA).

Using  the  far-IR extinction  maps  of  Schlegel  \etal (1998)  the  Galactic
foreground  extinction  in  the  direction  of  XRF  030528  is  estimated  as
E($B$--$V$)=0.62.  More  recent, several authors  have argued that  the far-IR
analysis  overestimates the  value of  E($B$--$V$) by  up to  $\sim$30\,\% for
fields at  low galactic lattitude. Dutra  \etal (2003) suggest  a rescaling of
the Schlegel \etal extinction by 0.75 which results in an E($B$--$V$)=0.46 for
the line of sight towards the  host of XRF 030528.  For the following analysis
we corrected the spectra according to  the rescaled value and consider this as
a lower limit of the foreground extinction.  All presented spectral parameters
(line fluxes \& luminosities) and absolute magnitudes have to be considered as
lower limits.

% --------------------------------------------------------------------------
\section{Results}

\begin{figure*}[t]
\centering \includegraphics[width=0.65\textwidth,angle=-90]{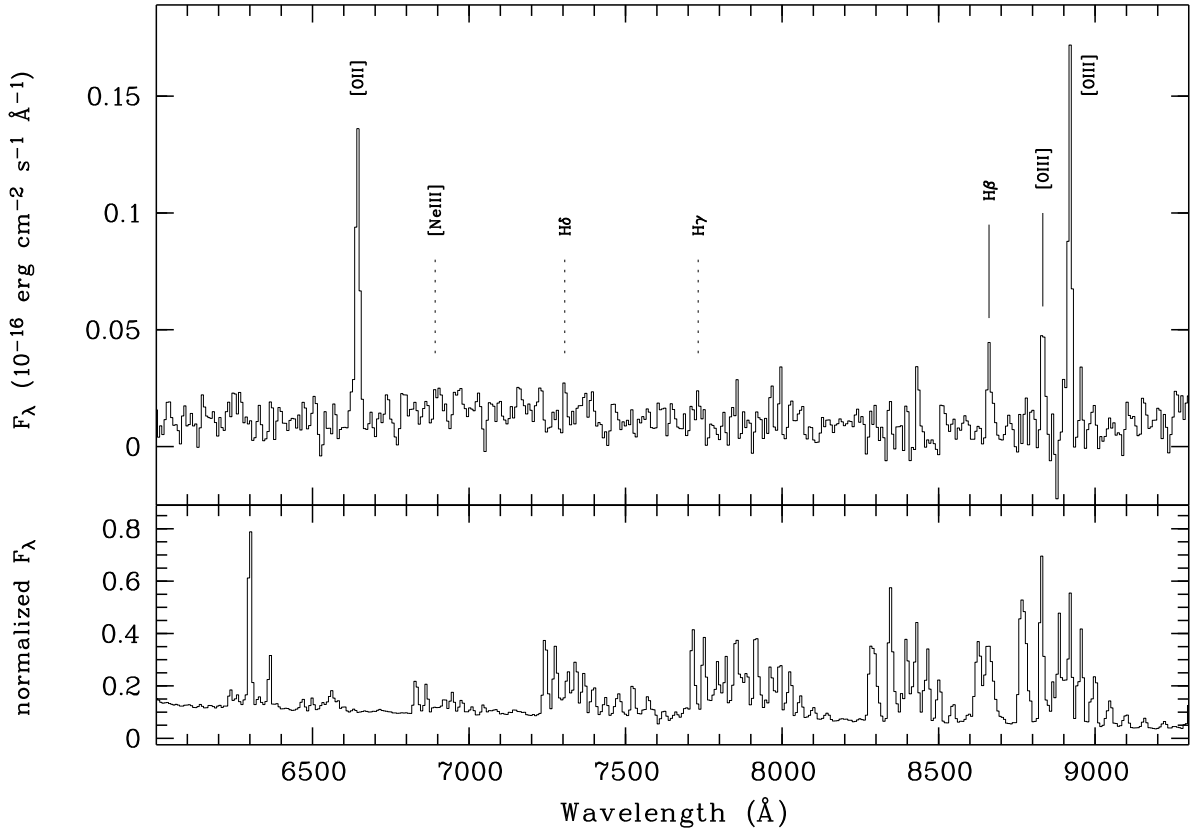}
\caption{{\it top:} Final FORS2 long-slit  spectrum of the host galaxy with an
  exposure  of $\sim$2\,hrs. The  positions of  the identified  emission lines
  (solid lines)  and a selection  of undetected lines (dashed)  are indicated.
  The host  continuum is only  marginally detected. Residuals of  sky emission
  features  are  visible  above   8000\,\AA.   {\it  bottom:}  Normalized  sky
  spectrum showing the prominent telluric emission lines.}
\label{fig:spectrum}
% n9bderedden
\end{figure*}

The   final   summed    spectrum   of   the   host   galaxy    is   shown   in
Figure~\ref{fig:spectrum}.   A  number   of  significant  emission  lines  are
detected   which  we  identified   as  [OII]   ($\lambda$3727\,\AA),  H$\beta$
($\lambda$4861\,\AA),      [OIII]     ($\lambda$4959\,\AA)      and     [OIII]
($\lambda$5007\,\AA)    at    a    helio-center    corrected    redshift    of
$z$=0.782$\pm$0.001.   This   corresponds   to   a  luminosity   distance   of
$D_L$=4.949$\pm$0.008\,Gpc using the  cosmological parameters given above.  At
this redshift H$\beta$  and the [OIII] emission lines  coincide with prominent
sky emission  features which complicates  the accurate extraction of  the line
properties and effects possible weak  detections of H$\gamma$ and H$\delta$ in
emission. Only [OII] falls into a wavelength range empty of sky features.

% --------------------------------------------------------------------------
% --------------------------------------------------------------------------

We measured  the emission  line strengths  using a gaussian  fit in  IRAF {\it
splot}.  The line fluxes are listed in Table~\ref{tab:lines} together with the
estimated redshifts  and the according  line luminosities.  The widths  of the
emission  lines  are consistent  with  the  instrumental  line width  (560\,km
s$^{-1}$ at $\lambda$=7000\AA) and do  not show signs of intrinsic broadening.
Consistent  with  the photometric  measurements  in  the  $V,R$ \&  $I$  bands
presented in R04, the host continuum is only marginally traced.  The resulting
uncertainty  in  the  continuum  flux  prevents  us  from  deriving  sensitive
equivalent widths for the lines.

%------------------------------------------------------------------
\begin{table}[h]
\begin{center}
\caption{Line    identifications.   Columns:   observer    frame   wavelengths
    $\lambda_{obs}$,  corresponding emission line  redshifts $z$,  line fluxes
    and  corresponding luminosities for  the lines  indicated in  the spectrum
    shown in Fig.~\ref{fig:spectrum}. Note that we do not apply a correction
    for underlying Balmer absorption.}
\begin{tabular}{lcccc}
\hline\hline
Line &  $\lambda_{obs}$ & $z$ & Flux & L \\
& \AA & & 10$^{-17}$ erg/s/cm$^{2}$ & 10$^{41}$ erg/s \\
\hline
$[OII]$ $\lambda$3727 & 6644 & 0.783 & 15$\pm$1 & 4.4$\pm$0.2\\
$[NeIII]$ $\lambda$3869 & 6891 & 0.781 & $<$1 & $<$0.3\\
$H$$_{\delta}$ $\lambda$4102 & 7306 & 0.781  & $<$2.5 & $<$0.7\\
$H$$_{\gamma}$ $\lambda$4340 & 7732 & 0.781 & $<$2.5 & $<$0.7\\
%$[OIII]$ $\lambda$4363 & 7775 & 0.782 & 1.0$\pm$0.2 & $<$0.3\\
%$HeII$ $\lambda$4686 & 8342 & 0.780 & 3.4$\pm$0.4 & 1.5$\pm$0.2\\
$H$$_{\beta}$ $\lambda$4861 & 8661 & 0.782 & 4.8$\pm$0.4 & 1.4$\pm$0.2\\
$[OIII]$ $\lambda$4959 & 8833 & 0.781 & 4$\pm$1 & 1.2$\pm$0.3\\
$[OIII]$ $\lambda$5007 & 8919 & 0.781 & 20$\pm$1 & 5.9$\pm$0.3\\
\hline\hline
\end{tabular}
\label{tab:lines}
\end{center}
\end{table}
%------------------------------------------------------------------

The  spectrum is  uncorrected for  possible intrinsic  extinction in  the host
galaxy. In order to get an estimate  on the extinction we used the Balmer line
ratios and fits to the  spectral energy distribution using the host photometry
presented in R04. Due to the shift of H$\alpha$ to the near-IR only the ratios
of  H$\gamma$/H$\beta$   and  H$\delta$/H$\beta$  can  be   used  to  estimate
E($B$--$V$). Comparing the line ratios derived from the strict upper limits of
the  line fluxes  for  H$\gamma$  and H$\delta$  with  the theoretical  values
(Brocklehurst  1971) we  find  that no  constrain  for the  extinction can  be
obtained.  In an independent attempt  we applied SED fitting to the broad-band
photometry in  $V,R,I,J_s,H$ \& $K$  using HyperZ (Bolzonella \etal  2000) and
theoretical spectral model templates of Bruzual \& Charlot (1993).  The sparse
photometric sampling of the  host together with the considerable uncertainties
provides  lower  and  upper  limits  of A$_V$=0  and  A$_V$=2.5,  respectively
($\chi^2$$<$1).  Thus, only a lower  limit of the intrinsic extinction A$_V$=0
can be derived.  Therefore,  the line strengths given in Table~\ref{tab:lines}
have to be considered as strict lower limits.
 
The unextincted star  formation rate (SFR) can be  derived from the luminosity
of the  [OII] emission  line, e.g.  using  the typical applied  calibration of
Kennicutt          (1998,          K98          hereafter),          SFR(\msun
yr$^{-1}$)=1.4$\pm$0.4$\times$10$^{-41}$L$_{[OII]}$.  Taking the measured line
luminosity  as  a  lower  limit  for  the  real  luminosity  we  estimate  the
unextincted star  formation to be  $>$6$\pm$2\,\msun yr$^{-1}$. A  measure for
the extinction corrected  SFR was proposed by Rosa-Gonzalez  \etal (2002, RG02
hereafter),   SFR(\msun   yr$^{-1}$)=8.4$\pm$0.4$\times$10$^{-41}$L$_{[OII]}$.
They derived ``unbiased''  SFR expressions, e.g. computed from  the [OII] line
luminosity and  the UV  continuum.  Correcting for  the effects  of underlying
stellar Balmer absorption their SFR  estimators bring into agreement the rates
measured with  different indicators, including the far-IR.   According to this
calibrator we derive an extinction corrected SFR of 37$\pm$4\,\msun yr$^{-1}$.

The UV  continuum probes  directly the emission  from young massive  stars and
thus  is another  measure  of the  unextincted  fraction of  the ongoing  star
formation.  The  optimal rest frame wavelength  (1500--2800\,\AA) lies outside
of our spectroscopic coverage  ($>$2950\,\AA\ at $z$=0.782), and the continuum
is only marginally  traced. Therefore, we used the best  fitting SED to derive
the flux at rest frame 2800\,\AA.  The resulting lower limit for the UV SFR is
4$\pm$1\,\msun   yr$^{-1}$  when   applying  the   K98   estimator,  SFR(\msun
yr$^{-1}$)=1.4$\pm$0.4$\times$10$^{-28}$L$_{\nu,UV}$,    and   17$\pm$3\,\msun
yr$^{-1}$  for  the corresponding  extinction  corrected  calibrator of  RG02,
SFR(\msun  yr$^{-1}$)=6.4$\pm$0.4$\times$10$^{-28}$L$_{\nu,UV}$.  For clarity,
the results are summarized  in Table~\ref{tab:sfr}. The derived star formation
rates are about  a factor of two lower than  the corresponding values obtained
using the [OII] emission line as indicator, which is fully consistent with the
spread  generally obtained  when using  various SFR  indicators  (e.g. Hopkins
\etal 2001).

%------------------------------------------------------------------
\begin{table}[h]
\begin{center}
\caption{Star formation rates and  specific star formation rates for different
  indicators (column  1: [OII] \&  UV continuum at 2800\,\AA)  and calibrators
  (column 2: Kennicut 1998 \&  Rosa-Gonzalez \etal 2002). The derived SFRs are
  given in  column 3 and  the columns 4  and 5 show  the specific SFRs  for an
  $L_{\star,B}$ galaxy and per unit solar mass, respectively.}
\begin{tabular}{ccccc}
\hline\hline
& & SFR & \multicolumn{2}{c}{specific SFR} \\
& & [\msun yr$^{-1}$] & [\msun yr$^{-1}$] & [\msun yr$^{-1}$  \msun$^{-1}$]\\
\hline
[OII] & K98 & 6$\pm$2 & 12$\pm$3 & 2$\times$10$^{-10}$\\
& RG02 & 37$\pm$4 & 74$\pm$6 & 1.2$\times$10$^{-9}$\\
\hline
UV & K98 & 4$\pm$1 & 8$\pm$2 &  1$\times$10$^{-10}$\\
& RG02 & 17$\pm$3 & 34$\pm$4 & 5$\times$10$^{-10}$\\
\hline\hline
\end{tabular}
\label{tab:sfr}
\end{center}
\end{table}
%------------------------------------------------------------------

The  metallicity of  the host  galaxy can  be derived  from the  emission line
indicator  R$_{23}$=log(([OII]+[OIII])/H$\beta$)  (Pagel  \etal  1979).   This
emipirical  indicator is  not unique  and typically  provides a  double branch
solution. This degeneracy can generally  be broken using other strong emission
lines (e.g.  [NII] $\lambda$6584\,\AA). For the host of XRF 030528 these lines
are not available but the value of R$_{23}$ falls onto the turnover of the two
branches.   Using  the  calibrations   compiled  in  McGaugh  (1991)  we  find
12+[log(O/H)]=7.7--8.5, which corresponds to a metallicity of 0.1--0.6 solar.

Knowing the redshift we determined the absolute magnitudes and luminosities of
the  host  galaxy in  various  photometric  bands.   The rest  frame  absolute
magnitudes (AB system, K-correction only) were derived using spectral template
fitting in HyperZ with the  intrinsic extinction fixed to zero.  The resulting
magnitudes are  shown in Table~\ref{tab:absMags} together  with the respective
magnitudes of an $L_\star$ galaxy in a Schechter distribution function and the
luminosities of the host in units of $L$/$L_\star$.

The host galaxy  is of the order of $L_\star$ in  the $U$-band and subluminous
at longer wavelengths, as expected  for an actively star forming galaxy.  Note
that the use of the Schlegel \etal extinction value (E($B$--$V$)=0.62) without
the correction  suggested by Dutra \etal  (see above) would  affect mainly the
bands shortwards of the rest frame $B$-band (roughly corresponding to observer
frame   $R$).   The  corresponding   luminosities  for   E($B$--$V$)=0.62  are
$L_U$$\sim$2.2$L_{\star,U}$ and $L_B$$\sim$0.9$L_{\star,B}$,  both a factor of
two brighter than the results for E($B$--$V$)=0.43.  Therefore, the magnitudes
and  luminosities given in  Table~\ref{tab:absMags} have  to be  understood as
lower limits.

%------------------------------------------------------------------
\begin{table}[h]
\begin{center}
\caption{Absolute magnitudes  of the host galaxy in  various photometric bands
(AB system) together with the absolute  magnitudes of an $L_\star$ galaxy in a
Schechter distribution  function and  the luminosity of  the host in  units of
$L$/$L_{\star}$.  Values for M$_{\star}$--5  log h$_{70}$ obtained for a range
in redshift consistent  with the redshift f our host  galaxy were adopted from
Dahlen  \etal (2005)  ($U,B,R_c$ \&  $J$) and  Cowie \etal  (1996)  ($K$). The
uncertainties in M  and $L$/$L_{\star}$ do not contain  the uncertainty in the
Galactic reddening (see Sect. 3)}

\begin{tabular}{cccc}
\hline\hline
rest frame band pass & M & M$_{\star}$ - 5 log h$_{70}$ & $L$/$L_{\star}$ \\
& [mag] & [mag] & \\
\hline
U & -20.5$\pm$0.1 & -20.3$\pm$0.1 & 1.2$\pm$0.2 \\
B & -20.7$\pm$0.1 & -21.4$\pm$0.1 & 0.5$\pm$0.1 \\
R$_c$ & -21.1$\pm$0.1 & -22.3$\pm$0.1 & 0.35$\pm$0.05 \\
J & -21.4$\pm$0.1 & -23.0$\pm$0.2 & 0.25$\pm$0.05 \\
K$_s$ & -21.6$\pm$0.1 & -23.5$\pm$0.2 & 0.17$\pm$0.05 \\ 
\hline\hline
\end{tabular}
\label{tab:absMags}
\end{center}
\end{table}
%------------------------------------------------------------------

In addition to the star formation  rate and luminosities we can also derive an
estimate  of the stellar  mass in  the host  galaxy applying  various methods.
Using  the  correlation between  the  mass and  rest  frame  $B$ and  $V$-band
magnitudes  of  Bell \etal  (2005),  assuming  a  Kroupa (2001)  initial  mass
function, we estimate the  mass to be $\sim$2$\times$10$^{10}$\,\msun.  As the
correlation  assumes a  wide  range in  stellar  ages the  obtained masses  of
galaxies with recent starburst activity, like  the host of XRF 030528, will be
overestimated. Nevertheless,  the obtained  mass is of  the same order  as the
stellar mass derived  from the rest frame $K$-band  mass-to-light ratio of 0.8
(Brinchmann \& Ellis 2000) of $\sim$9$\times$10$^{9}$\,\msun.

Using  the luminosity  of the  host  and its  stellar mass  the specific  star
formation rate in units of \msun yr$^{-1}$ for an $L_{\star,B}$ galaxy and per
unit mass  can be estimated.  The  specific star formation rates  for the RG02
SFR calibrators are 74$\pm$6 and 34$\pm$4\,\msun yr$^{-1}$ for an $L_{\star,B}$
galaxy   derived  from  the   [OII]  and   UV  estimators,   respectively  and
1.2$\times$10$^{-9}$  and  5$\times$10$^{-10}$\,\msun yr$^{-1}$  \msun$^{-1}$,
respectively (see also Tab.~\ref{tab:sfr}).

% --------------------------------------------------------------------------
\section{Discussion}

\begin{figure}[h]
\centering \includegraphics[width=0.355\textwidth,angle=-90]{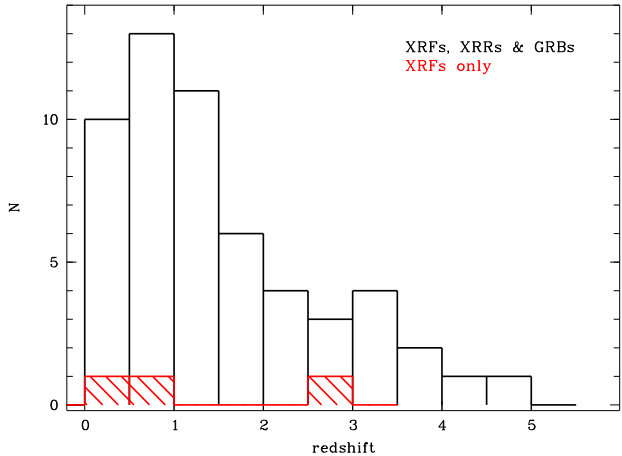}
\caption{Distribution of accurately measured redshifts for XRFs, XRRs and GRBs
  together  (empty  histogram) and  XRFs  only  (hatched).   The total  sample
  contains  55 bursts  with GRB  050803 (Bloom  \etal 2005)  being  the latest
  entry. The XRFs are 020903, 030528  and 030429 at $z$=0.251, 0.782 and 2.65,
  respectively.}
\label{fig:zDistr}
\end{figure}

Accurate distance  measurements for extragalactic  high-energy transients like
XRFs,  XRRs  and GRBs  are  important  to  discriminate between  the  proposed
theoretical  models and  for the  application  of the  events as  cosmological
probes.   While  the redshift  distribution  of  classical  GRBs contains  now
$\sim$50 events  covering a redshift  range from $z$=0.0085 to  $z$=4.511, the
distance scale of  XRFs is much less determined.   The here presented redshift
measurement of XRF 030528 gives the third profound distance to an XRF obtained
so far  and shows that  XRFs, classified according  to the {\it  HETE-2} X-ray
over  $\gamma$-ray  fluence  ratio  in  the  observer  frame,  cover  a  range
consistent with that of their classical brothers (Fig.~\ref{fig:zDistr}).

% --------------------------------------------------------------------------
% --------------------------------------------------------------------------

A possible  scenario associates XRFs with  GRBs at high  redshift (Heise \etal
2001). The  three obtained XRF distances  indicated that this can  not be true
for at  least some of  the events. Nevertheless,  the shift of the  rest frame
peak energy to  lower values in the observer frame can  not be neglected.  for
XRF 030528  the observer  frame peak energy  of $E^{obs}_{peak}$=32$\pm$5\,keV
corresponds to $E^{rest}_{peak}$=57$\pm$9\,keV at a redshift of $z$=0.782.  We
modeled the  high energy  spectrum of  the burst using  a Band  function (Band
\etal  1993) with  parameter values  given for  XRF 030528  in  Sakamoto \etal
(2005) and  classified it according  to the  {\it HETE-2}  scheme in  the rest
frame.  The  obtained value of  log($S_X$/$S_\gamma$)$\sim$--0.17 demonstrates
that XRF  030528 would be  defined as  an X-ray rich  burst at its  rest frame
rather than an XRF.

A similar  estimate was  obtained for XRF  030429 ($z$=2.65).  For  this event
$E^{obs}_{peak}$=35$^{+12}_{-8}$\,keV   (Sakamoto   \etal   2005)  shifts   to
$E^{rest}_{peak}$=128$^{+43}_{-30}$\,keV in the  rest frame.  This corresponds
to log($S_X$/$S_\gamma$)$\sim$--0.43 which places the burst at the boarderline
between  ``classical'' GRBs  and XRRs.   From the  three XRFs  with accurately
known   redshift   only   for   XRF  020903   ($E^{rest}_{peak}$=6\,keV)   its
classification remains that  of an XRF also in the  rest frame.  However, this
event was  especially soft  and the physical  similarity with bursts  like XRF
030528 or 030429 is uncertain.

As shown above,  the identification of bursts based  on observer frame fluence
ratios  can in  most  cases only  be  seen as  an operational  classification.
Naturally, a more  sophisticated seperation would require to  be based on rest
frame rather  than observer frame  properties.  Nevertheless, it  becomes more
and more  evident that XRFs, XRRs and  GRBs form a continuum  of objects which
generally questions the necessity of  such a classification of events (both in
the observer frame  as well as in  the rest frame) and evaluates  it as purely
operational.

Knowing the redshift, the isotropic  equivalent energy for the prompt emission
of       XRF       030528       can       be      determined       to       be
$E_{iso,\gamma}$=2.0$\pm$0.7$\times$10$^{52}$\,erg in the 2--400\,keV observer
frame energy range. Together  with $E^{rest}_{peak}$=57\,keV, XRF 030528 falls
at the lower end of  the correlation of $E_{iso,\gamma}$ and $E^{rest}_{peak}$
found  by Amati  \etal  (2002).   Assuming the  validity  of the  relationship
between  $E^{rest}_{peak}$   and  the  collimation   corrected  total  energy,
E$_\gamma$, proposed  by Ghirlanda \etal  (2004), the collimation  fraction of
the  burst can be  estimated.  Equation  (3) of  Ghirlanda \etal  (2004) gives
E$_\gamma$=4.8$\pm0.7$$\times$10$^{49}$  for  $E^{rest}_{peak}$=57\,keV.  This
implies a collimation  factor of 2.4$\times$10$^{-3}$ and an  opening angle of
4$\degs$, which is in the typical range for GRB jets (Frail \etal 2001).

\bigskip

The principal aim of the observation  presented here was the estimation of the
redshift  of XRF 030528  and thus  the further  establishment of  the distance
scale of XRFs.  As it turned out, XRF 030528 is no longer an XRF but an XRR in
the rest  frame. The good quality  of the spectroscopic data  provided us with
the possibility to study also the underlying host galaxy in more detail.  From
the  rest frame  UV  continuum and  the  [OII] emission  line luminosity  star
formation rate  values ranging  from 4 to  37\,\msun yr$^{-1}$  were obtained.
Despite the large  uncertainty of the estimate, we can  conclude that the host
galaxy exhibits a significant level of ongoing star formation, similar to what
was found by Christensen \etal (2004) in a sample of 10 GRB host galaxies.

XRF  030528 occurred in  a galaxy  which appears  sub-luminous in  the near-IR
($\sim$0.2$L_{\star,K}$)       and        rest-frame       optical       bands
($\sim$0.5$L_{\star,B}$). This is also  typical for the hosts of long-duration
GRBS  studied so  far (e.g.,  Sokolov \etal  2001; le  Floc'h \etal  2003; 
Christensen \etal 2004).  Furthermore, the gas  in the host was found to be of
sub-solar  metallicity  (0.1--0.6\,$z$)  using  the  emission  line  indicator
R$_{23}$, in agreement with a recent  study of three low-$z$ GRB host galaxies
by Sollerman \etal  (2005).  All this suggests that the host  of XRF 030528 is
indeed an actively star forming galaxy as emphasized in R04.

Only a  small number  of estimates for  the stellar  mass content in  GRB host
galaxies have  been performed  so far (e.g.   Sokolov \etal 2001;  Chary \etal
2002).  For  the host  of XRF 030528  we obtained  the stellar mass  using two
independent  indicators applied to  the absolute  magnitudes derived  from the
photometry  presented in  R04.   Both  methods gave  a  consistent results  of
$M$$\sim$10$^{10}$\,\msun, which  places the host  among the most  massive GRB
hosts  at a  similar  redshift. Note  that  the previous  mass estimates  were
obtained using synthesis model fitting while we used indicators based directly
on the absolute $B$ \& $V$ and $K$-band magnitudes.

At  a  redshift   of  $z$=0.782,  the  angular  extent   of  the  host  galaxy
($\sim$1\farcs5; R04) corresponds to  a linear size of $\sim$11\,kpc (ignoring
possible inclination  effects).  This indicates that  the host is  not a dwarf
galaxy but more comparable in size with a spiral galaxy like the Milky Way.

The comparison of the specific star formation rate and the stellar mass of the
host galaxy  with galaxies  in the  FORS Deep Field  and GOODS-South  field at
similar redshift,  shows that it falls  into the group of  young actively star
forming galaxies (Feulner \etal 2005).   This indicates once more (e.g., Bloom
\etal 2003;  Soderberg \etal 2004; Fynbo  \etal 2004) that  XRFs are similarly
associated  with  star forming  regions  in the  universe  as  GRBs and  again
suggests  the familiarity  between  these events.   Unfortunately, the  sparse
afterglow and host photometry prevents  us from deriving an accurate constrain
on a  supernova associated with  XRF 030528. The obtained  $K$-band brightness
14.8\,days post-burst  provides an upper  limit for the absolute  magnitude of
any super nova  of $K$=--21.6\,mag, around 2\,mag brighter  than most luminous
core-collaps super nova observed in the near-IR (Mannucci \etal 2003.

While  the indications  that XRFs,  XRRs  and GRBs  form a  continuum in  many
properties are more and more evident,  the origin of the soft spectral peak in
XRFs is not revealed for all events.  In the case of XRF 030528 we showed that
the classfication in the observer frame differs from that obtained in the rest
frame. At least in XRF 020903, the  only ``real'' XRF with a known distance so
far, this is  not true. In order to distinguish  between the individual models
and to estimate  the rest frame properties of  the events accurately, redshift
measurements  for  XRFs  are  still  very important.  These  require  accurate
localizations which  will be provided by  the ongoing {\it  HETE-2} mission as
well  as,  although  with  a  lower  rate, by  the  {\it  Swift}  instruments.
Nevertheless,  the detection  of  very soft  events  like XRF  020903 at  high
redshift (z$>$1) will require sensitive instruments at even lower energies.

% --------------------------------------------------------------------------
\bigskip

\noindent
{\it Acknowledgments.}
This  work  is  based  on  observations collected  at  the  European  Southern
Observatory,  Chile,  under  proposal  075.D-0539.  We  thank  D.~Hartmann  \&
S.~Savaglio for the fruitful discussions. We thank the anonymous referee for
valuable comments.

\bigskip

% --------------------------------------------------------------------------


\begin{thebibliography}{}
\bibitem{} Amati, L., Frontera, F., Tavani, M., \etal 2002, A\&A, 390, 8
\bibitem{} Band, D., Matteson, J., Ford, L. \etal 1993, ApJ, 413, 281
\bibitem{} Barraud, C., Olive, J.-F., Lestrade, J.P., \etal 2003, A\&A, 400, 1021
\bibitem{} Barraud, C., Daigne, F., Mochkovitch, R., \& Atteia, J.-L., 2005,
  A\&A in press, astro-ph/0507173
\bibitem{} Bell, E.F., Papovich, C., Wolf., C., \etal 2005, ApJ, 625, 23
\bibitem{} Bloom, J.S., Fox, D., van Dokkum, P.G., \etal 2003, ApJ, 599, 957
\bibitem{} Bloom, J.S., Perley, D., Foley R., \etal 2005, GCN, 3758
\bibitem{} Bolzonella, M., Miralles, J.-M., Pello, R., 2000, A\&A, 363, 476
\bibitem{} Brinchmann, J., \& Ellis, R.S., 2000, ApJ, 536, L77
\bibitem{} Brocklehurst, M., 1971, MNRAS, 153, 471
\bibitem{} Bruzual, G. \& Charlot, S. 1993, ApJ, 405, 538
\bibitem{} Butler, N., Dullighan, A., Ford, P., \etal 2003, GCN, 2279
\bibitem{} Butler, N., Dullighan, A., Ford, P., \etal 2004, in Proc. Gamma-Ray
  Bursts: 30 Years of Discovery: Gamma-Ray Burst Symposium,  ed. E.E. Fenimore
  \& M. Galassi, AIPC Vol 727, p.435
\bibitem{} Chary, R., Becklin, E.E., \& Armus, L., 2002, ApJ, 566, 229
\bibitem{} Christensen, L., Hjorth, J., \& Gorosabel, J., A\&A, 425, 913
\bibitem{} Cowie, L.L., Songaila A., Hu, E.M., \& Cohen, J.G., 1996, AJ, 112, 389
\bibitem{} Dahlen, T., Mobasher, B., Somerville, R.S., \etal 2005, ApJ
  accepted, astro-ph/0505297
\bibitem{} Dermer, C.D., Chiang, J., \& B\"ottcher, M., 1999, ApJ, 513, 656
\bibitem{} Dutra, C.M., Ahumada, A.V., Claria, J.J., Bica, E., \& Barbuy, B.,
  2003, A\&A, 408, 287
\bibitem{} Feulner, G., Gabasch, A., Salvato, M., \etal 2005, ApJL, submitted
\bibitem{} Fox, D.B., 2004, GCN, 2630
\bibitem{} Frail, D.A., Kulkarni, S.R., Sari, R. \etal 2001, ApJ, 562, L55
\bibitem{} Fynbo, J.P.U., Sollerman, J., Hjorth, J., \etal 2004, ApJ, 609, 962
\bibitem{} Ghirlanda, G., Ghisellini, G., \& Lazzati, D., 2004, ApJ, 616, 331 
\bibitem{} Greiner, J., Rau, A., \& Klose, S., 2003a, GCN, 2271
\bibitem{} Harrison, F.A., Yost, S., Fox., D.B., \etal 2001, GCN 1143
\bibitem{} Heise, J.,  in't Zand, J, Kippen, R.M., \& Woods P.M., 2001, in
  Gamma-Ray Bursts in the Afterglow Era, ed. E. Costa, F. Frontera, \&
  J. Hjorth (Berlin: Springer), 16
\bibitem{} Hjorth, J., Sollerman, J., M{\o}ller, P., \etal, 2003, Nature, 423, 847
\bibitem{} Hopkins, A.M., Connolly, A.J., Haarsma, D.B., \&  Cram, L.E., 2001, AJ, 122, 288
\bibitem{} Huang Y.F., Dai, Z.G., \& Lu, T., 2002, MNRAS, 332, 735
\bibitem{} Kelson, D.D., Koviak, K., Berger, E., \& Fox, D.B., 2004, GCN, 2627
\bibitem{} Kennicutt, R.C., 1998, ARA\&A, 36, 189
\bibitem{} Lamb, D.Q., Donaghy, T.Q., \& Graziani, C., 2005, ApJ, 620, 355
\bibitem{} le Floc'h, E., Duc ,P-A., Mirabel, I.F., \etal 2003, A\&A, 400, 499
\bibitem{} Mannucci, F., Maiolino, R., Cresci, G., \etal 2003, A\&A, 401, 519
\bibitem{} McGaugh, S.S., 1991, ApJ, 380, 140
\bibitem{} Pagel, B.E.J., Edmunds, M.G., Blackwell, D.E., \etal, 1979, MNRAS,
  255, 325
\bibitem{} Preece, R.D., Briggs, M.A., Mallozzi, R.S., \etal 2000, ApJS, 126, 19
\bibitem{} Rau, A., Greiner, J., Klose, S., \etal 2004, A\&A, 427, 815
\bibitem{} Rosa-Gonzalez, D., Terlevich, E. \& Terlevich, R., 2002, MNRAS,
  332, 283
\bibitem{} Sakamoto, T., Lamb, D.Q., Graziani, C., \etal 2004, ApJ, 602, 875
\bibitem{} Sakamoto, T., Lamb, D.Q., Kawai, N., \etal 2005, ApJ, 629, 311
\bibitem{} Schlegel, D.J., Finkbeiner D.P., \& Davis, M. 1998,
ApJ, 500, 525
\bibitem{} Soderberg, A.M., \etal 2003, AAS Meeting 203, 132.07
\bibitem{} Soderberg, A.M., Kulkarni, S.R., Berger, E., \etal 2004,  ApJ, 606, 994
\bibitem{} Soderberg, A.M., Kulkarni, S.R., Fox, D.B., \etal 2005, ApJ, 627, 877
\bibitem{} Sollerman, J., \"Ostlin G., J.P.U. Fynbo, \etal 2005, astro-ph/0506686
\bibitem{} Sokolov, V.V., Fathkhullin, T.A., Castro-Tirado, A.J., \etal 2001,
  A\&A, 372, 438
\bibitem{} Stanek, K.Z., Matheson, T., Garnavich, P. M., \etal, 2003, ApJ,
  591, L17
\bibitem{} Taylor, G.B., Frail, D.A., \& Kulkarni, S.R., 2001, GCN, 1136
\bibitem{} van Dokkum, P.G., \& Bloom, J.S., 2003, GCN, 2380
\bibitem{} Weidinger, M., Fynbo, J.P.U., Hjorth, J., \etal 2003, GCN, 2215
\bibitem{} Yamazaki, R., Ioka, K., \& Nakamura, T., 2002, ApJ, 571, L31
\bibitem{} Zeh, A., Klose, S., \& Hartmann, D.H., ApJ, 2004, 609, 952
\end{thebibliography}
\end{document}